\documentclass[11pt,a4paper]{article}
\usepackage[hyperref]{acl2017}
\usepackage{times}
\usepackage{latexsym}
\usepackage{subfigure}
\usepackage{url}
\aclfinalcopy
\def\aclpaperid{21}

\usepackage{graphicx} 
\usepackage{subfigure} 

\usepackage{algorithm}
\usepackage{algorithmic}

\def\Degsym{D}

\def\transpos#1{#1^{\top}}
\def\s#1{{\cal #1}}
\def\norme#1{\left\|#1\right\|}

\usepackage{amsthm}
\usepackage{amssymb,amsmath,amsthm,amsfonts}

\theoremstyle{definition}

\title{Seating Assignment Using Constrained Signed Spectral Clustering}

\author{
Jo\~{a}o Sedoc \\
        Computer \& Information Science\\
        University of Pennsylvania \\
\texttt{joao@cis.upenn.edu}\\
        \And
Aline Normoyle\\
        Computer \& Information Science\\
        University of Pennsylvania \\
        \texttt{alinen@seas.upenn.edu}
}

\date{}

\begin{document}

\maketitle

\begin{abstract}
In this paper, we present a novel method for constrained cluster size signed spectral clustering (CSS) which allows us to subdivide large groups of people based on their relationships. In general, signed clustering only requires $K$ hard clusters and does not constrain the cluster sizes. We extend signed clustering to include cluster size constraints. Using an example of seating assignment, we efficiently find groups of people with high social affinity while mitigating awkward social interaction between people who dislike each other.
\end{abstract}

\label{intro}
\section{Introduction}

Despite being a commonplace aspect of life, reasoning about relationships is difficult. 
For a compelling example, consider the emotionally
high-stakes task of wedding planning, and in particular: seating charts. 
Personal histories, backgrounds, and personalities of a large number
of people have to match to avoid hurt feelings, discordant conversations, and perhaps even a decades-long grudge. 

In this paper, we introduce a novel method for 
constrained signed spectral clustering (CSS) which allows us to represent and analyze complex 
sets of relationships. CSS clustering builds on the idea of modeling social affinity
(i.e. ``like'' or ``dislike'') as a graph (\citet{cartwright1956structural}) to
support cluster size constraints. Using this technique, a user can express
relationships pairwise, or as cliques where all people like each other. 

The CSS clustering algorithm tackles the problem of assigned seating by maximizing the amount of positive relationships between people at the tables, while minimizing the negative ones. CSS clustering aims to do this over all tables simultaneously. As with most clustering problems, we first find the solution to a quickly solvable continuous problem, and then rotate to the best integer solution. CSS clustering is an extension of multiclass signed clustering~\citep{jeanfullspectral}. 
Our main innovation is using the continuous solution as a ranked input for the National Resident Matching Program (NRMP) algorithm~\citep{roth1999redesign}.
\begin{figure}[h!]
    \centering
    \includegraphics[width=0.5\textwidth]{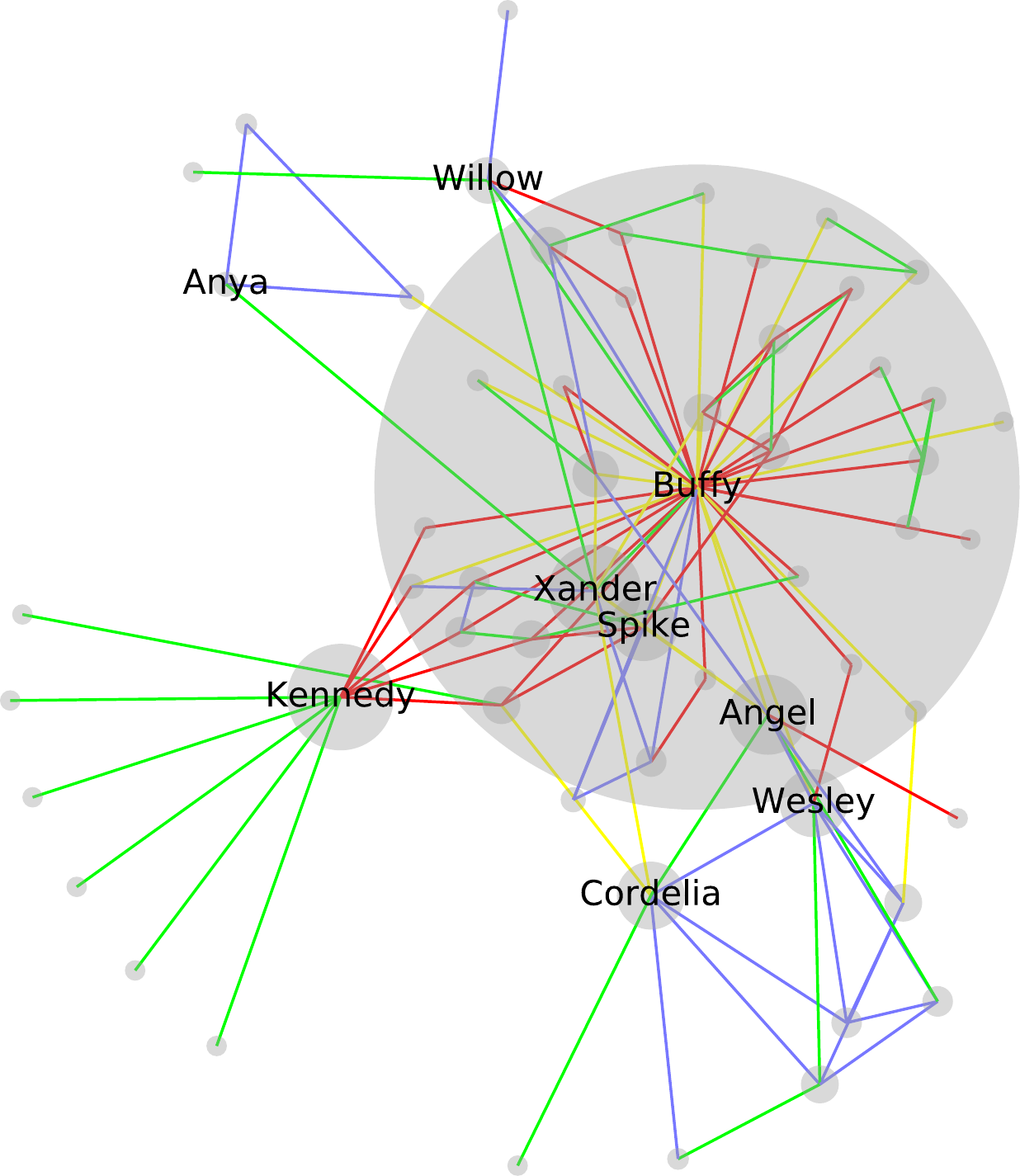}
    \caption{Social graph centered around fictional character, Buffy Summers. Green lines indicate strong positive affinity; Blue lines positive affinity; yellow lines negative affinity; and red lines strong negative affinity.}
    \label{fig:socialnetwork}
\end{figure}
Throughout this work we focus on the concrete example of assigned seating to ground this problem; however, the method is extensible to a wider range of problems.

\section{Seating Assignment Problem}
\label{seating}
The assigned seating problem is a graph problem where we need to assign $N$ people to $K$ tables. Each table $i$ has a maximum number of seats $T^i$. The people are represented by vertices, $v_i$, and the affinity (i.e. ``like'' and ``dislike'') between person $i$ and person $j$ is represented by weighted edges, $w_{ij}$. The affinity of one person with herself is set to zero, $w_{ii}=0$. Now, we formalize our graph $G$ as $G = (V,W)$ where $V$ is the group of people and $W$ is the matrix of their pairwise affinities. 
We define the degree of a vertex $v_i$ to be the sum of the absolute values of its edge weights, so 
$
d(v_i) = \sum_{j=0}^{N}{|w_{ij}|}
$
with the corresponding signed degree matrix $\overline{D} = diag(d(v_1), \dots, d(v_N))$. The degree of a person is a measure of that person's social connectivity. The signed graph Laplacian is $\overline{L} = \overline{D} - W$.

Having defined the graph $G$ and its intrinsic qualities, we needed to decide how to group people into tables. We partition the set of vertices $V$ into clusters, $(C_1, \dots, C_K)$, where $K$ is the total number of clusters. Each cluster $C_j$ has at least one vertex $v_i$, but no more than $T^j$ vertices. 

\label{method}
\section{Optimization Problem}
In this section we further formalize the SSC clustering optimization problem. 
A formulation of the multi-class signed spectral clustering (signed normalized cut) problem is stated as, 
\begin{align*}
& \mathrm{minimize}     &  \sum_{j = 1}^K 
\frac{\transpos{(X^j)} \overline{L} X^j}{\transpos{(X^j)}\overline{\Degsym} X^j}& &  &  &\\
& \mathrm{subject\ to} & 
 \transpos{(X^i)} \overline{\Degsym} X^j = 0,  & &  &  &\\
& & \quad 1\leq i, j \leq K,\> 
i\not= j,  & &  X\in \s{X}, & &  & 
\end{align*}
where $X$ is the $N\times K$ matrix whose $j$th column is $X^j$~\citep{jeanfullspectral}.
Note that $X^j$ corresponds to $C_j$ where $x^j_i = 1$ if and only if $v_i \in C_j$. 
If we let
\begin{align*}
\s{X}  = \Big\{[X^1\> \ldots \> X^K] \mid
X^j = (x_1^j, \ldots, x_N^j) , \\
\>
x_i^j \in \{1, 0\}, \> X^j \not= 0
\Big\}
\end{align*}
our solution set is
\begin{align*}
\s{K}  = \Big\{
X  \in\s{X}  \mid \transpos{(X^i)}& \overline{\Degsym} X^j = 0,\\ 
& 1 \leq i, j \leq K, \quad i\not= j
\Big\}.
\end{align*}
CSS clustering constrains cluster sizes such that, 
\begin{align*}
& \mathrm{minimize}     &  \sum_{j = 1}^K 
\frac{\transpos{(X^j)} \overline{L} X^j}{\transpos{(X^j)}\overline{\Degsym} X^j}& &  &  &\\
& \mathrm{subject\ to} & 
 \transpos{(X^i)} \overline{\Degsym} X^j = 0,  & &  &  &\\
& & \transpos{({X^i})}\mathbf{1}_N \leq T^i,  & &  & &  & \\
& & \quad 1\leq i, j \leq K,\> 
i\not= j,  & &  X\in \s{X}, & &  & 
\end{align*}
where  $\mathbf{1}_N$ is the vector of ones of length $N$. 
As stated, this optimization is an integer quadratic programming problem, which is known to be NP-hard~\citep{wolsey1998integer}. As a result, we relaxed the problem by removing the constraints that $X  \in\s{X}$ and $\transpos{({X^i})}\mathbf{1}_N \leq T^i$. After finding the solution to the relaxed continuous problem, then the closest integer solution is found. 

\subsection{Relaxed Problem }

As in \citet{jeanfullspectral}, we find the relaxed eigenvalue problem, %
\begin{align*}
& \mathrm{minimize}     &  &  
\mathrm{tr}(\transpos{Y}\overline{\Degsym}^{-1/2} \overline{L} \overline{\Degsym}^{-1/2} Y)& &  &  &\\
& \mathrm{subject\ to} &  & 
\transpos{Y} Y = I. 
 & &  & &  
\end{align*}
\noindent The minimum of the relaxed problem, $Z$, 
is achieved by the $K$ unit eigenvectors associated with the smallest eigenvalues of the signed normalized Laplacian, $L_{\mathrm{sym}}= \overline{D}^{-1/2} \overline{L} \overline{D}^{-1/2}$. 

\subsection{Finding the Approximate Discrete Solution}

As initially proposed by \citet{yu2003multiclass} and extended by \citet{jeanfullspectral}, $Z$ is rotated and scaled to the closest orthogonal solution using an invertible diagonal scaling matrix, $\Lambda$, and distance preserving transformations $R \in \mathbf{O}(K)$ (see section 4.5 of \citet{jeanfullspectral} for more detail.) This is equivalent to minimizing $\norme{X - ZR\Lambda}_F$, where $\norme{A}_F$ is the Frobenius norm of $A$. After finding the unconstrained solution for signed clustering $X^*$, and also, the probabilistic solution $X^{**}$ by zeroing out the negative entries of $X$ and subsequently normalizing the rows to sum to one, the NRMP algorithm is applied to reassign vertices from clusters which have too many vertices.

\begin{algorithm}[H]
   \caption[CSSC Algo]{Constrained Signed Spectral Clustering}
   \label{alg:cssc_algo}
\begin{algorithmic}[1]
{\small
   \STATE {\bfseries Input:} Weight matrix $W$ (without isolated nodes), number of clusters $K$, maximum cluster sizes $T^i$ and termination threshold $\epsilon$.
   \STATE Using the degree matrix $\overline{D}$, and the signed Laplacian $\overline{L}$, compute the signed normalized Laplacian $\overline{L}_{sym}$.

   \STATE Initialize $\Lambda = I$,
   ${X} = \overline{D}^{- \frac{1}{2}}U$ 
   where $U$ is the matrix of the eigenvectors corresponding to the K smallest eigenvalues of $\overline{L}_{sym}$.~\footnote{We use fast randomized SVD\citep{halko2011finding}. While in general one cannot use fast SVD for finding the smallest eigenvalues, in our case we know that the eigenvalues are bounded.} 
   \WHILE{$\norme{X - ZR\Lambda}_F > \epsilon$ }
   \STATE Minimize $\norme{X - ZR\Lambda}_F$ with respect to $X$ holding $Z$, $R$, and $\Lambda$
fixed.
   \STATE Fix $X$, $Z$, and $\Lambda$,  find $R\in \mathbf{O}(K)$
that minimizes  $\norme{X - ZR\Lambda}_F$.
   \STATE Fix $X$, $Z$, and $R$,  find a 
diagonal invertible matrix $\Lambda$ that
minimizes  $\norme{X - ZR\Lambda}_F$.
    \ENDWHILE
    
   \STATE Find the unconstrained discrete solution $X^*$ by choosing the largest entry $x_{ij}$ on row $i$ set $x_{ij} = 1$ and all other $x_{ij} = 0$ for row $i$. 
   \STATE Find $X^{**}$ by setting all $x_{ij}<0$ to $0$, and subsequently renormalizing $x_{ij} = \frac{x_{ij}}{\sum_{l=1}^{m}{x_{lj}}}$. 
   \STATE For each cluster $C_i$ if $\transpos{(X^i)}\mathbf{1} > T^i$ then remove $x_{ij}$ with minimum value and add row $i$ from $X^{**}$ to the stack $A$ and set $x_{ij}=0$.
   \STATE For each cluster $C_i$ and vertex $v_l \in A$ where $\transpos{(X^i)}\mathbf{1} < T^i$ compute $r_{il} = \sum_{i\in C_i}{w_{ij}}$.
   \STATE Run the NRMP algorithm using $A$ and $R$ and assign the vertices in $A$ to clusters back to $X^*$.
   \STATE {\bfseries Output:} $X^*$.
   }
\end{algorithmic}
\end{algorithm}
\vspace{-0.2in}

After running algorithm~\ref{alg:cssc_algo}, the isolated vertices (i.e. people share edges with no one) are randomly assigned to tables which are not full. Steps 10-13 deviate from the original algorithm~\citep{sedoc2016semantic}. This is the main theoretical innovation of the paper.

\section{Results}
\label{results}

To demonstrate the approach, we run CSS clustering on a 
mid-sized group of 58 fictional characters, taken from the 
show \emph{Buffy the Vampire Slayer}, with the goal of creating a drama-minimizing seating arrangement. This show, which 
features a high school cheerleader whose destiny is to 
kill vampires, contains a large cast of characters with rich backstories, making it challenging to organize links of enemies, friends, rivalries, grudges, and x-boyfriends (Figure~\ref{fig:socialnetwork} gives a sense of the complexity between characters' connections). We use Buffy as a light-hearted, but non-trivial, example 
of one of the most frustrating and tedious tasks of event planning. 

\begin{table*}[ht]
    \centering
\begin{tabular}{l|llllllllll}
Table ID & 0 & 1 & 2 & 3 & 4 & 5 & 6 & 7 & 8 & 9 \\
\hline
\# seated & 10 & 5 & 5 & 9 & 4 & 6 & 2 & 3 & 4 & 10\\
volume & 44.1 & 22.6 & 32.5 & 43.2 & 10.5 & 51.0 & 10 & 10.2 &30.3 & 60.1 \\
\# components &  7 & 2  & 1  & 5 & 3 & 1 & 1 & 2  & 1 & 1 \\
\hline 
\end{tabular}
    \caption{Solved seating arrangement. We ran our algorithm for a room containing 10 tables with a maximum of 10 seats each. The volume is the sum of the edges between each table member. For example, table 6 has two people with strong affinity and the volume is 10. Lastly, we compute the number of connected components for each table. 
    For example, tables 0 and 3, which have many disconnected components correspond to villain tables -- most have few friends among the other guests and strong animosity from the other tables.
    }
    \label{tab:results}
\end{table*}

\subsection{Specifying constraints}

\begin{figure}
    \centering
    \includegraphics[width=0.5\textwidth]{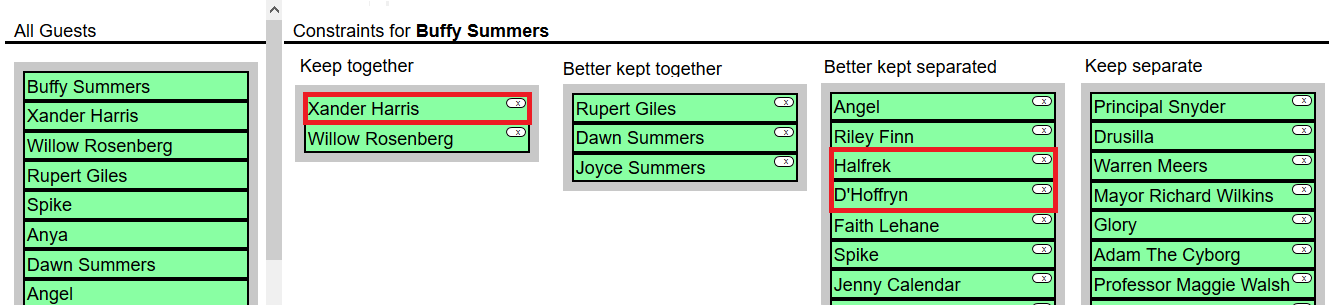}
    \caption{User interface for specifying constraints. Users can specify four choices of relationship: \emph{keep together} (strong positive affinity), \emph{better kept together} (positive affinity), \emph{better kept separated} (negative affinity), and \emph{keep separate} (strong negative affinity). Above, Buffy wants to sit with her best friend Xander but doesn't like his girlfriend's demon friends, Halfrek and D'Hoffryn. 
    }
    \label{fig:ui}
\end{figure}

To help users specify constraints, we have built a javascript 
user interface that reads in a CSV file of names and allows users to drag and drop people based on their affinity (Figure~\ref{fig:ui}). Users can specify four choices of relationship: \emph{keep together}, \emph{better kept together}, \emph{better kept separated}, and \emph{keep separate}. \emph{Keep together} and \emph{Keep separate} represent important relationships -- encoded with graph weights of 10 and -10 respectively. The remaining constraints are encoded with 1 and -1. If no constraints are specified between people, we assume they have a neutral affinity and weight them with a slight positive value of 0.1. All affinities are symmetric (e.g. we do not allow for A to like B, but B hates A). If contradictory affinities are specified (e.g. if A must sit with B and C, but B hates C), the user is warned.

To see why seating arrangements quickly become difficult to manage, consider the main character, Buffer Summers. Her closest friends, Willow Rosenburg and Xander Harris, clearly should sit with her. Analogously, her numerous enemies should clearly be seated apart. However, once tertiary relationships begin to be considered, decisions become much harder. For example, Xander might like to sit with his deceased girlfriend Anya. But Anya is friends with vengeance demons Halfrek and D'Hoffryn, who Buffy doesn't like. Using the application, the user can simply specify the relationship constraints and let the algorithm handle managing them.

\subsection{Solving for seating arrangements}

\begin{figure}
    \centering
    \includegraphics[width=0.4\textwidth]{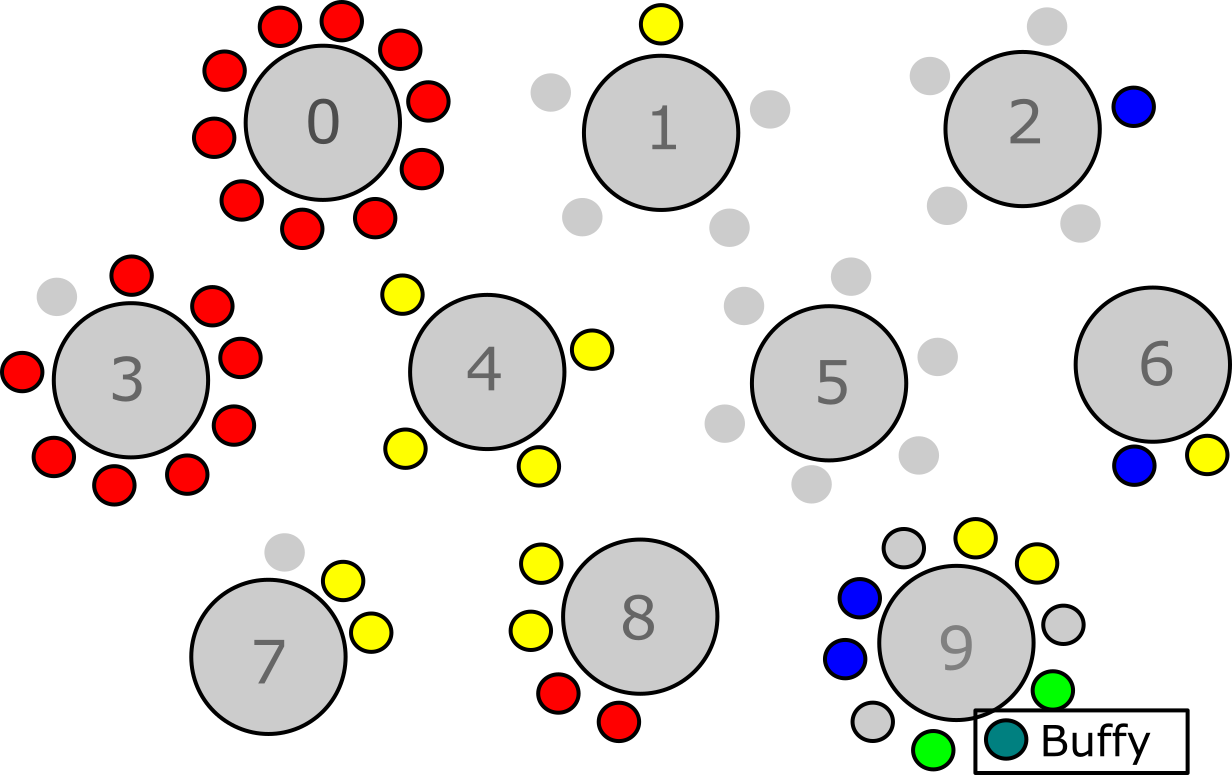}
    \caption{Locations of all characters with a relationship to Buffy. Green seats represent strong affinity; Blue, positive; Yellow, negative; Red, strongly negative; Gray, neutral. Buffy is seated next to her closest friends while enemies sit at other tables.}\label{fig:buffyTable}
\end{figure}

The CSS clustering algorithm satisfies the seating constraints described in the previous section. In Figure~\ref{fig:buffyTable}, we see that Buffy is seated with Xander and Willow. Furthermore, 
both Xander and Willow have been seated with their dates, Anya and Tara. Although Buffy would rather not sit with Halfrek and D'Hoffrin, 
Anya's presence for them outweighed her dislike. In a real situation, the user could override this decision. Importantly, none 
of Buffy's enemies are at her table. The villains tend to be grouped together because most of the other guests dislike them and few villains have affinities with each other (as demonstrated by the larger numbers of disconnected components for tables 0 and 3. See Table~\ref{tab:results}).

\begin{figure}
    \centering
        \includegraphics[width=0.2\textwidth]{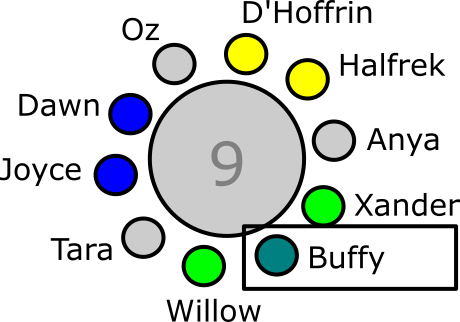}
    ~ 
        \includegraphics[width=0.2\textwidth]{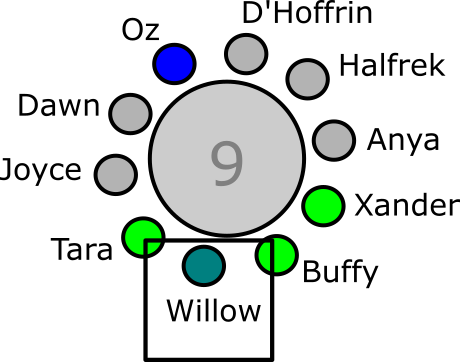}
    ~ 
        \includegraphics[width=0.2\textwidth]{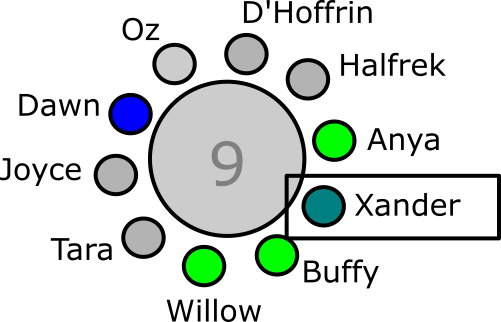}
        \includegraphics[width=0.2\textwidth]{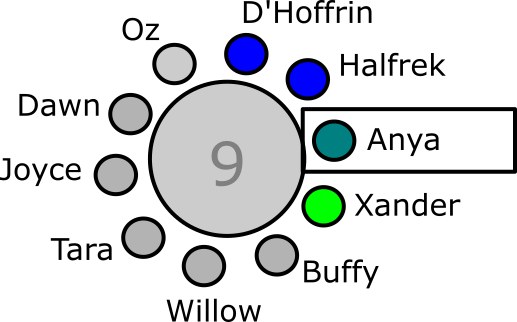}
    \caption{Relationships between members of table 9. We color the seats of table 9 based on the perspectives of Buffy, Willow, Xander, and Anya. Green seats represent strong affinity; Blue, positive; Yellow, negative; Red, strongly negative; Gray, neutral.}\label{fig:table9}
\end{figure}


\section{Discussion}
To our knowledge there is no well-founded algorithm for seating assignment. 
\citet{tang2016survey} present an excellent survey of signed networks for social media.
However, the constrained group size problem has been ignored.
Other methods which incorporate negative edge weight such as must-link/cannot-link spectral clustering~\citep{rangapuramconstrained}, k-Oppositive Cohesive Groups \citep{chu2016finding}, complete positive factorization~\citep{zass2005unifying} do not constrain cluster size.
There is a constrained formulation using fixed size Markov Cluster Process~\citep{van2001graph}; however, this is not generalized for signed graphs. 

The process of entering social affinities is still quite tedious; however, social affinities can be predicted using e-mail correspondence, social media. This could extend existing work on predicting social relationships~\citep{iyyer2016feuding}.

For multiclass spectral clustering, the Cheeger inequality gives a bound on how far away the approximate solution can be from the exact solution~\citep{chung1997spectral,lee2014multiway}. We leave it for future work to bound the approximate solution for CSS clustering.

\section{Conclusion}
\label{conclusion}
Constrained signed spectral clustering is an important enhancement to signed spectral clustering using normalized cuts. We have shown that in the example of an assigned seating arrangement CSS clustering can provide practical solutions.

\bibliography{assigned_seating}
\bibliographystyle{acl_natbib}

\end{document}